\documentclass{article}

\usepackage{microtype}
\usepackage{hyperref}
\usepackage{booktabs}

\usepackage[preprint]{mlsys2026}
\hypersetup{
  pdftitle={Coordination-Free Workload Regularization for Variable-Cost Distributed Training},
  pdfauthor={Tao Huang, Chen Zhu, Antonio Alonso, Daming Li, Adarsh Goura, Jaideep Ray, Yitong Zhou},
  pdfsubject={Preprint}
}

\mlsystitlerunning{BCL Workload Regularization}

\begin{document}

\twocolumn[
\mlsystitle{Coordination-Free Workload Regularization for Variable-Cost Distributed Training}

\begin{mlsysauthorlist}
\mlsysauthor{Tao Huang}{linkedin}\textsuperscript{$\dagger$}
\mlsysauthor{Chen Zhu}{linkedin}
\mlsysauthor{Antonio Alonso}{linkedin}
\mlsysauthor{Daming Li}{linkedin}
\mlsysauthor{Adarsh Goura}{linkedin,utdallas}\textsuperscript{$\ast$}
\mlsysauthor{Jaideep Ray}{linkedin}
\mlsysauthor{Yitong Zhou}{linkedin}
\end{mlsysauthorlist}

\mlsysaffiliation{linkedin}{LinkedIn Corporation, Sunnyvale, CA, USA}
\mlsysaffiliation{utdallas}{The University of Texas at Dallas, Richardson, TX, USA}
\centerline{\small\textsuperscript{1}\,LinkedIn Corporation, Sunnyvale, CA, USA}
\centerline{\small\textsuperscript{2}\,The University of Texas at Dallas, Richardson, TX, USA}
\mlsyskeywords{machine learning systems, distributed training, variable-length workloads, scheduling}

\vskip 0.3in

\begin{abstract}
Synchronous data-parallel training usually assigns every rank the same number of records, even
when record costs vary by orders of magnitude. For variable-length models, this converts input
variation into idle time because each step waits for its most expensive rank-local batch. We
present Batch Cost Lookahead (BCL), a coordination-free data-loading policy that buffers a bounded
rank-local window and repartitions it into fixed-cardinality packs with balanced estimated cost.
BCL preserves the trainer's record-count contract and requires neither cross-rank metadata
exchange nor runtime modification. In simulation over public workload distributions, enabling
BCL reduces per-rank estimated batch-cost variability by up to 82.8\%.
Across 13 NeMo-RL SFT corpora on eight H200 GPUs, rank-local BCL
runs 1.050--1.167$\times$ faster than pooled MFFD.
Compiled H200 measurements on an internal sequence workload show 1.197--1.343$\times$
speedup over packed execution from one to four nodes.
We also identify a memory-feasibility boundary: a p99 cap for
high-variance or OOM-prone data bounds outliers before BCL balances the rest.
These results show that substantial
variable-cost imbalance can be removed at the data-loader boundary, while clarifying when data
preparation or broader coordination is still necessary.
\end{abstract}
]

\chead{\small\bf BCL Workload Regularization}
\begingroup
\renewcommand{\thefootnote}{$\dagger$}
\footnotetext{\raggedright Corresponding author: Tao Huang (\texttt{taohuang@linkedin.com}).}
\endgroup
\begingroup
\renewcommand{\thefootnote}{$\ast$}
\footnotetext{\raggedright Adarsh Goura was a summer 2026 intern at LinkedIn.}
\endgroup

\section{Introduction}

In synchronous data-parallel training, each rank processes a local batch, computes gradients,
and participates in collective synchronization \citep{li2020pytorch,goyal2017accurate}.
Workload imbalance is pronounced for variable-length sequences. The cost of token-wise operations
grows with sequence length, while dense self-attention includes a quadratic sequence-length
component \citep{vaswani2017attention,dao2022flashattention}. Consequently, two batches with the
same number of records can have very different costs when one contains longer sequences. In
synchronous training, this imbalance lies on the critical path because each step waits for the
slowest rank \citep{chen2016revisiting}. The resulting idle time tends to grow with workload
variability and the number of participating ranks \citep{dean2013tail,chen2016revisiting}.

Sequence packing and variable-length kernels reduce padding work within a batch
\citep{krell2021packing,dao2022flashattention}, but they do not determine how records should be
grouped across rank-local batches in a synchronized step. Cross-worker balancing can use global
metadata to make that decision, but recent designs require additional communication or runtime
coordination \citep{wang2026libra,wan2026odc}. This paper asks a narrower systems question:

\begin{quote}
\emph{How much variable-cost imbalance can be removed using only bounded information already
available to each rank-local data loader?}
\end{quote}

We study Batch Cost Lookahead (BCL), a local workload-regularization policy. A worker buffers
$K$ future fixed-cardinality batches, estimates a cost for each record, and partitions the
$K B$ records into $K$ packs of exactly $B$ records. The objective is to reduce the maximum pack
cost while preserving the existing fixed-cardinality interface. Each input reader schedules independently, without exchanging record costs with other
readers or changing the model and distributed collectives.

Our production trainer expects a fixed record count, so BCL retains that interface.
Token-budget batching can improve utilization, but varying the number of records
per step requires careful loss normalization \citep{ott2018scaling,krell2021packing}.
BCL regroups the same records within a bounded window while emitting exactly
$B$ records per pack. This makes BCL complementary to padding elimination
and packed execution: packing removes wasted work within a pack, while BCL reduces cost
dispersion among packs that must synchronize.
Fixed-token packing can also leave unequal attention work across ranks
\citep{wang2026libra}; globally coordinated assignment offers another response to this skew
\citep{li2026hydraulis}.

Our evaluation separates three forms of evidence. First, a self-contained simulation over six
checked-in length arrays derived from public datasets isolates the scheduling effect. With batch
size 128, eight ranks,
and a quadratic capped-length cost, increasing lookahead from $K=1$ to $K=32$ reduces per-rank
estimated batch-cost variability by up to 82.8\%.
Second, NeMo-RL SFT runs \citep{nemorl} compare independent per-rank
input with pooled MFFD. At $B=8$ on eight
H200 GPUs, BCL improves wall-clock step time by 1.050--1.167$\times$ across
13 corpora; this includes both scheduling and input-path differences.
Third, compiled H200
measurements on an internal Feed sequential-ranking workload whose model setting is described
publicly by \citet{hertel2026feed} compare BCL at $K=32$ with packed execution from one to four
nodes. The measured speedup ranges from 1.197$\times$ to 1.343$\times$.

BCL cannot split an overlong record, so we set the length policy
before comparing packers. Earlier feasibility probes found five corpora that passed
uncapped smoke tests, four that used p99 proactively for high length variability,
and four for which p99 enabled runs that had OOMed uncapped. Every NeMo comparison
arm uses the same per-corpus cap, separating the effect of grouping from
changes in truncation.

This paper makes four contributions:
\begin{itemize}
  \item We formulate a coordination-free operating point for variable-cost balancing: rank-local
  scheduling under a fixed-cardinality and bounded-reordering data-loader contract.
  \item We implement fixed-cardinality balancing within bounded reader windows.
  \item We evaluate BCL in a public-distribution simulation, NeMo-RL SFT on 13
  corpora, and internal H200 runs on one to four nodes.
  \item We identify regimes in which local scheduling is insufficient and length capping or
  broader coordination is required.
\end{itemize}

\paragraph{Paper organization.}
Section~2 motivates the cost model and distinguishes cross-rank imbalance from padding waste.
Section~3 describes BCL and its fixed-cardinality implementation. Section~4 develops the
reducible and irreducible regimes. Sections~5--7 present the public-distribution, NeMo, and
multi-node internal evaluations; each introduces its workload, setup, reporting provenance, and
results together. Section~8 situates BCL against related work, Section~9 summarizes
deployment boundaries, and Section~10 concludes.

\section{Motivation and Cost Model}
\label{sec:cost-model}

\subsection{Fixed cardinality is not fixed work}

Consider synchronous data-parallel training with $R$ ranks and a prescribed per-rank batch
cardinality $B$. At step $t$, every rank processes exactly $B$ records before participating in
the same synchronization barrier. This fixed-count contract controls how many records contribute
to an optimizer step, but it does not control how much computation those records require.

For variable-length data, the record lengths assigned to a rank determine its workload. Two
ranks can therefore receive the same number of records while one processes substantially more
tokens or more attention work. Because the synchronized step waits for the slowest rank, this
cost difference becomes idle time on the faster ranks
\citep{li2020pytorch,chen2016revisiting}. The problem addressed in this paper is not unequal
record count; it is unequal work under an intentionally equal-record-count interface.
This interface describes our production streaming use case; token-budget
and packed batches with variable record counts can exhibit the same synchronization imbalance.
The formal metrics here use fixed $B$; Section~\ref{sec:nemo-evaluation} compares
token-budget packing using measured wall-clock time.

\subsection{Padding waste and synchronization imbalance}

Padding and cross-rank imbalance are separate inefficiencies. In a padded batch, shorter
sequences execute against the longest sequence's shape. Sequence packing and variable-length
kernels reduce this intra-batch waste by avoiding computation on padding
\citep{krell2021packing,dao2022flashattention}. They do not, however, guarantee that the useful
work assigned to each rank is equal. A packed rank containing several long records can still take
longer than another packed rank containing the same number of shorter records.

BCL targets this residual synchronization imbalance. It is complementary to packed execution:
packing changes how the records inside a pack are represented, whereas BCL changes which records
are grouped into packs that will execute at the same synchronization boundary. The evaluation
therefore reports packed baselines separately rather than attributing padding removal to BCL.

\subsection{Per-record and per-pack cost}

Let $\ell$ denote a record's sequence length and let $\ell_{\max}$ be an optional feasibility cap.
We use the effective length $\bar{\ell}=\min(\ell,\ell_{\max})$ and the first-order proxy
\begin{equation}
  c(\ell)=\alpha\bar{\ell}+\beta\bar{\ell}^{2},
  \label{eq:record-cost}
\end{equation}
where $\alpha,\beta\geq0$. The linear term represents token-wise work such as projections and
feed-forward layers, while the quadratic term represents dense self-attention
\citep{vaswani2017attention,dao2022flashattention}. Setting $(\alpha,\beta)=(1,0)$ gives the
linear, or L1, mode; $(0,1)$ gives the quadratic, or L2, mode; and nonzero coefficients give a
coefficient-weighted blend.

For rank $r$ at step $t$, the modeled pack cost is additive:
\begin{equation}
  C_{r,t}=\sum_{i=1}^{B} c(\ell_{r,t,i}).
  \label{eq:pack-cost}
\end{equation}
Only relative costs matter to the scheduler, so multiplying every cost by the same positive
constant does not change an assignment or a normalized balance metric. The cap and coefficients
are workload policy, not properties inferred by BCL. The public-distribution simulation uses
capped L2 cost to isolate the scheduling effect; each measured evaluation states its length
policy and controls alongside its results.

The cost estimate in Equation~\ref{eq:pack-cost} guides how BCL groups records, but GPU step
time also depends on kernels, communication, data loading, and compilation. We measure
step time on hardware rather than treating improved estimated balance as a speedup.

\subsection{Critical-path metrics}

The modeled critical-path cost of synchronized step $t$ is
\begin{equation}
  C^{\max}_{t}=\max_{1\leq r\leq R} C_{r,t}.
  \label{eq:critical-cost}
\end{equation}
We report two complementary normalized metrics. Pack-cost coefficient of variation measures the
relative temporal spread of emitted pack costs:
\begin{equation}
  \mathrm{CV}_{\mathrm{pack}}
  =\frac{1}{R}\sum_{r=1}^{R}
   \frac{\mathrm{std}_{t}(C_{r,t})}{\mathrm{mean}_{t}(C_{r,t})}.
  \label{eq:pack-cv}
\end{equation}
A lower value means each rank receives more consistent modeled work over time. The synchronization
straggler ratio measures how much modeled work lies on the cross-rank critical path:
\begin{equation}
  \rho
  =\frac{\mathrm{mean}_{t}(C^{\max}_{t})}
         {\mathrm{mean}_{r,t}(C_{r,t})}.
  \label{eq:straggler}
\end{equation}
Here $\rho=1$ is perfect modeled balance, while $\rho=1.10$ means the per-step
bottleneck pack is, on average, about 10\% costlier than the average pack.
The excess over perfect modeled balance
is $\rho-1$; only part of it may be removed by rank-local scheduling.

These metrics answer different questions. Pack-cost variability measures regularity within each
rank's emitted stream; the straggler ratio measures the synchronized penalty across ranks. BCL's
fixed-cardinality objective, developed next, reduces the maximum pack cost within each bounded
local window without changing the records consumed or the number of records per step.

\section{BCL Design and Implementation}
\label{sec:bcl-design}

\subsection{Rank-local deployment boundary}

BCL runs independently inside each rank's data-loader path. Let $B$ be the required number of
records per emitted batch and $K$ the lookahead in batches. A worker stages a bounded window of
\begin{equation}
  W=K B
  \label{eq:window-size}
\end{equation}
records from its own stream, assigns those records to $K$ output packs, and then resumes filling
the next window. It reads no other rank's stream, exchanges no per-record metadata, and moves no
records across ranks.

The deployment contract has three invariants. First, every staged record is emitted exactly once.
Second, every output pack contains exactly $B$ records. Third, reordering is confined to the
current $W$-record window. BCL therefore preserves the optimizer's record-count cadence and the
existing distributed collectives while trading a bounded amount of host-side staging and
reordering for better balance. Length capping, rendering, and tokenization occur before
scheduling; BCL never silently truncates an already-rendered record.

\subsection{Fixed-cardinality objective}

For one window, let $P_1,\ldots,P_K$ be a disjoint partition of the record indices, with
$|P_k|=B$ for every pack. Using the cost function from Equation~\ref{eq:record-cost}, the
fixed-cardinality objective is
\begin{equation}
  \min_{P_1,\ldots,P_K}
  \max_{1\leq k\leq K}\sum_{i\in P_k} c(\ell_i).
  \label{eq:bcl-objective}
\end{equation}
This is a cardinality-constrained makespan objective: it directly targets the most expensive pack
that may become a synchronized straggler. Lower variance is a useful consequence, but not the
optimization target.

The decision version contains strongly NP-complete 3-PARTITION when packs have three
records and their number can vary \citep{garey1979computers}; dynamic programming
applies only to restricted cases
with few packs and bounded integer costs. BCL starts with a longest-processing-time-first
ordering, assigning each record to the least-loaded pack that is not full
\citep{graham1969bounds}. Classical LPT bounds do not cover exact cardinality and local
refinement, so we evaluate balance empirically rather than claiming an approximation bound.

\subsection{Bounded scheduling procedure}

Algorithm~\ref{alg:bcl} gives the fixed-cardinality procedure. A deterministic per-record tie key
makes the assignment reproducible. A scan of the non-full packs selects the
least-loaded pack while enforcing cardinality. Refinement considers exchanges
between the current heaviest and lightest packs and accepts one only when it
strictly reduces the maximum pack cost.
Because refinement uses exchanges rather than one-way moves, it cannot change pack cardinality.

\begin{algorithm}[t]
\caption{Fixed-cardinality BCL on one rank-local window}
\label{alg:bcl}
\begin{algorithmic}[1]
\REQUIRE Lengths $\ell_1,\ldots,\ell_W$ with $W=KB$; cost $c$; tie key $\tau$; rounds $Q$
\STATE $I \leftarrow$ sort indices by $(-c(\ell_i),\tau(i))$
\STATE Initialize $K$ empty packs $P_k$ with loads $L_k\leftarrow0$
\FOR{each $i$ in $I$}
  \STATE $k \leftarrow$ least-loaded pack with $|P_k|<B$, breaking ties by index
  \STATE $P_k\leftarrow P_k\cup\{i\}$; $L_k\leftarrow L_k+c(\ell_i)$
\ENDFOR
\FOR{$q=1,\ldots,Q$}
  \WHILE{a heavy--light exchange strictly lowers $\max_k L_k$}
    \STATE Select a deterministic improving exchange
    \STATE Apply the exchange and update the two affected loads
  \ENDWHILE
\ENDFOR
\STATE \textbf{return} $P_1,\ldots,P_K$
\end{algorithmic}
\end{algorithm}

The construction conserves records and cardinality by design.

\subsection{Streaming integration}

The public simulation, NeMo input path, and multi-node loader share
the fixed-cardinality scheduling contract. In the NeMo run, each GPU worker reads
only its own shards and schedules $K B$ records into $K$ packs of exactly $B$.
The worker consumes its own packs; neither records nor cost metadata are exchanged
between ranks.

For the pooled reference, we use NeMo-RL's modified first-fit
decreasing (MFFD) packer \citep{johnson1985mffd,nemorl} from the earlier global-input
campaign. It repacks the current $R B$ documents into $R$ token-budget rows;
per-rank BCL instead looks ahead across batches without seeing other ranks'
records. Because the loaders differ, the measured speedup reflects both
packing and input-pipeline costs.

\subsection{Cost and systems overhead}

For $W=KB$ records, stable sorting costs $O(W\log W)$ and scanning the $K$ packs
for each assignment costs $O(WK)$. Two refinement rounds repeat improving
heavy--light exchanges until none remains; storage is $O(W)$ for
the staged records, costs, and assignments. The scheduler operates on compact cost metadata after
rendering, so it adds no GPU collective and does not increase the optimizer batch or the number
of active records in a training step.

Larger $K$ stages more records on the host, delays the first emitted pack, and permits
farther reordering. It cannot make an infeasible record fit or remove kernel and
communication work shared by all ranks.

\section{Reducibility and Coordination Boundaries}
\label{sec:boundaries}

\subsection{A single-record lower bound}

Scheduling can only rearrange the records available in a window; it cannot split a record or
make that record cheaper. For one rank-local window, let
\begin{equation}
  S=\sum_{i=1}^{W} c(\ell_i),\qquad
  \bar{C}=\frac{S}{K},\qquad
  m=\max_i c(\ell_i),
  \label{eq:window-summary}
\end{equation}
where $\bar{C}$ is the average cost of a $B$-record pack. Any valid partition has
\begin{equation}
  C_{\max}^{\star}\geq\max(\bar{C},m).
  \label{eq:item-lower-bound}
\end{equation}
The first term is the average-load bound; the second follows because the largest record must
appear in some pack. Normalizing by $\bar{C}$ gives the single-record irreducibility index
\begin{equation}
  \eta=\frac{m}{\bar{C}}=\frac{K m}{S}.
  \label{eq:irreducibility}
\end{equation}
When $\eta\ll1$, no individual record forces a large residual and the window is potentially
reducible by assignment. When $\eta\geq1$, one record alone costs at least as much as an average
pack, so even an optimal scheduler has a normalized maximum of at least $\eta$.

This diagnostic is a lower bound, not a complete attainability test. Multiple large records,
exact-cardinality constraints, and a small window can leave additional discrete imbalance even
when $\eta<1$. Conversely, increasing $K$ can expose more complementary records, but it cannot
lower the floor imposed by an unchanged $m$.
A p99 cap can reduce the dominant record cost $m$. If the capped window
has $\eta<1$ and its packs fit in memory, the single-record bound no longer forces a
straggler; BCL can then reduce the assignment-induced part of the critical path.

\subsection{Why bounded lookahead helps}

Let the per-record cost random variable $X=c(\ell)$ have mean $\mu$, standard deviation $\sigma$,
and coefficient of variation $\mathrm{CV}_X=\sigma/\mu$. Under an independent random-assignment
approximation, a $B$-record pack has mean $B\mu$ and coefficient of variation
\begin{equation}
  \mathrm{CV}_{B}\approx\frac{\mathrm{CV}_X}{\sqrt{B}}.
  \label{eq:random-pack-cv}
\end{equation}
If $R$ rank-local pack costs have similar distributions, an order-statistics approximation gives
\begin{equation}
  \rho_{\mathrm{random}}
  \approx 1+a_R\frac{\mathrm{CV}_X}{\sqrt{B}},
  \label{eq:random-straggler}
\end{equation}
where $a_R$ is the standardized expected maximum and increases with the number of synchronized
ranks \citep{dean2013tail}. This explains why the same workload variation creates more
straggler opportunity at larger scale.

If record costs are independent and each rank can divide its $KB$ records into $K$
nearly equal-cost packs of $B$, those packs approach their rank-window mean. The
relative spread of that mean scales as $\mathrm{CV}_X/\sqrt{KB}$, suggesting
\begin{equation}
  \rho_{\mathrm{BCL}}
  \sim 1+a_R\frac{\mathrm{CV}_X}{\sqrt{KB}}.
  \label{eq:bcl-scaling}
\end{equation}
  Reusing $a_R$ for rank-window means is an approximation; its value depends on
  the distribution of the costs being maximized.
  The relation suggests diminishing gains as $K$ grows. Correlated
streams and scheduling-dependent pack costs can change that trend. Section~5 tests
modeled balance; Sections~6--7 measure step time on hardware.

\subsection{The rank-divergence floor}

Local lookahead also cannot change the total cost already assigned to a rank. For a block of $K$
synchronized steps, let rank $r$ receive window total $S_r$ and average pack cost
$\bar{C}_r=S_r/K$. Define the cross-rank mean
\begin{equation}
  \bar{C}_{\mathrm{all}}=\frac{1}{R}\sum_{r=1}^{R}\bar{C}_r
\end{equation}
and the rank-divergence index
\begin{equation}
  \delta=\frac{\max_r\bar{C}_r}{\bar{C}_{\mathrm{all}}}.
  \label{eq:rank-divergence}
\end{equation}
For aligned steps with one fixed window per GPU rank, local permutations cannot reduce
the block's normalized average critical path below $\delta$: they cannot move cost
between rank windows.

When ranks receive shuffled shards of the same distribution, $\delta$ is typically close to one
and temporal lookahead can address most assignment-induced variation. When shards or user
populations differ systematically, $\delta$ grows and the missing information is genuinely
cross-rank. A global scheduler may then improve balance by exchanging metadata or moving samples,
at the cost of a larger communication domain and tighter runtime coupling
\citep{wang2026libra,wan2026odc}.

\subsection{Choosing the required intervention}

Table~\ref{tab:boundary-actions} connects the single-record and rank-divergence
bounds with the remaining modeled straggler excess. Memory feasibility is checked
before scheduling. Both $\delta$ and $\rho$ are at least one; 1.10 represents a
10\% modeled excess over the corresponding average.

\begin{table}[H]
\caption{Workload boundary and appropriate intervention.}
\label{tab:boundary-actions}
\centering
\small
\begin{tabular}{@{}p{0.25\columnwidth}p{0.68\columnwidth}@{}}
\toprule
Diagnostic & Implication \\
\midrule
$\eta<1$, $\delta\approx1$ & Local variation is potentially reducible; increase bounded
lookahead until marginal gains saturate. \\
$\eta\geq1$ & A record meets or exceeds mean modeled pack cost; cap, split,
or change its representation before scheduling. \\
$\delta>1$ & Rank-window means differ; if material, reshard or use cross-rank
coordination. \\
$\rho\approx1$ & Little modeled imbalance remains; limited modeled
scheduling headroom. \\
\bottomrule
\end{tabular}
\par\smallskip
\parbox{\columnwidth}{\footnotesize\textit{Notation:} $\eta$ compares the largest
record cost with the mean pack cost in one window; $\delta$ compares the heaviest
rank-window mean with the mean across ranks; $\rho$ (Eq.~\ref{eq:straggler})
compares the average synchronized maximum pack cost with the average pack
cost. Equations~\ref{eq:random-straggler} and~\ref{eq:bcl-scaling}
approximate $\rho$ under random assignment and idealized BCL, respectively.}
\end{table}

Memory infeasibility is the strongest form of the single-record boundary. If one rendered record
cannot fit within the available memory, no assignment can make it feasible. The length policy
must therefore be established before BCL. Earlier uncapped probes distinguish proactive caps from caps prompted
by an OOM (Section~\ref{sec:discussion}); the streaming comparison fixes p99 caps
across arms. Earlier $B=8$ no-lookahead runs on CodeParrot and OpenWebMath OOMed on
oversized rows; per-rank BCL completes on both. Scheduling cannot make an individually
infeasible record fit.

\section{Public-Distribution Scheduling Simulation}
\label{sec:public-simulation}

\subsection{Question and public inputs}

Using the same sampled records in both arms, we test whether BCL steadies modeled pack
costs and lowers the synchronized maximum. GPU step time is measured separately.

The six checked-in arrays contain 165,011 integer token lengths derived from Stanford
Alpaca~\citep{taori2023alpaca}, the train split of the CodeParrot
clean-valid dataset~\citep{codeparrot}, Dolly~\citep{conover2023dolly},
OpenHermes~2.5~\citep{teknium2023openhermes},
OpenWebMath~\citep{paster2023openwebmath}, and the HuggingFaceH4 UltraChat~200k
subset~\citep{ding2023ultrachat}. They contain no source text. The public source identifier,
rendering rule, sample count, and SHA-256 checksum for every array are retained with the
experiment. Table~\ref{tab:public-inputs} reports the uncapped length statistics.

\begin{table}[H]
\caption{Public length arrays used by the scheduling simulation. Percentiles are token lengths
before the 1,024-token cost cap.}
\label{tab:public-inputs}
\centering
\small
\begin{tabular}{lrrr}
\toprule
Dataset & Records & p50 & p99 \\
\midrule
Alpaca & 30,000 & 59 & 270 \\
CodeParrot & 30,000 & 975 & 20,423 \\
Dolly & 15,011 & 101 & 1,146 \\
OpenHermes & 30,000 & 237 & 2,468 \\
OpenWebMath & 30,000 & 763 & 16,900 \\
UltraChat & 30,000 & 1,051 & 3,004 \\
\bottomrule
\end{tabular}
\end{table}

\subsection{Paired simulation design}

For each dataset, the benchmark draws one cost stream for each of $R=8$ ranks and reuses that
exact stream for both scheduling arms. Each arm therefore consumes the same sampled input stream on every rank;
only window-local grouping and emitted order change. We draw each rank's records independently, with
replacement. The results describe modeled workload balance, not measured GPU step time.
The run contains 1,000 synchronized steps with per-rank cardinality
$B=128$, seed zero, and capped quadratic cost
\begin{equation}
  c(\ell)=\min(\ell,1024)^2.
\end{equation}
The BCL-disabled arm emits arrival-order packs with $K=1$. The BCL arm uses $K=32$ with
cardinality-constrained LPT initialization and two iterative refinement rounds. Because 1,000 is not
divisible by 32, the last eight steps use a smaller eight-pack window rather than drawing or
discarding additional records. Constructed pack order is deterministically shuffled within each
rank window before ranks are paired into synchronized steps.

The primary metrics are the pack-cost coefficient of variation and straggler ratio defined in
Section~\ref{sec:cost-model}. We report their relative reductions as
\begin{equation}
  \Delta_{\mathrm{CV}}
  =1-\frac{\mathrm{CV}_{K=32}}{\mathrm{CV}_{K=1}},
\end{equation}
and
\begin{equation}
  \Delta_{\rho}
  =1-\frac{\rho_{K=32}-1}{\rho_{K=1}-1}.
\end{equation}
The second expression measures the fraction of modeled straggler \emph{excess} removed; it does
not treat the unavoidable unit baseline as removable work.

\subsection{Results}

Table~\ref{tab:public-results} shows the paired result. BCL reduces pack-cost variability by
81.4--82.8\% on every retained distribution. The modeled straggler ratio falls from
1.050--1.345 under arrival-order batching to 1.009--1.066 with BCL, removing 81.0--83.9\% of
modeled straggler excess.

\begin{table}[H]
\caption{Paired public-distribution simulation. Both arms consume the exact same sampled record
stream on every rank. $\Delta_{\rho}$ is the reduction in $\rho-1$.}
\label{tab:public-results}
\centering
\small
\begin{tabular}{lrrr}
\toprule
Dataset & CV, $K=1$ & CV, $K=32$ & $\Delta_{\mathrm{CV}}$ \\
\midrule
Alpaca & 0.187 & 0.032 & 82.8\% \\
CodeParrot & 0.051 & 0.009 & 81.5\% \\
Dolly & 0.228 & 0.041 & 82.1\% \\
OpenHermes & 0.133 & 0.024 & 82.2\% \\
OpenWebMath & 0.066 & 0.012 & 81.4\% \\
UltraChat & 0.036 & 0.006 & 82.3\% \\
\bottomrule
\end{tabular}

\vspace{0.4em}

\begin{tabular}{lrrr}
\toprule
Dataset & $\rho$, $K=1$ & $\rho$, $K=32$ & $\Delta_{\rho}$ \\
\midrule
Alpaca & 1.300 & 1.048 & 83.9\% \\
CodeParrot & 1.072 & 1.014 & 81.2\% \\
Dolly & 1.345 & 1.066 & 81.0\% \\
OpenHermes & 1.195 & 1.036 & 81.4\% \\
OpenWebMath & 1.094 & 1.017 & 82.4\% \\
UltraChat & 1.050 & 1.009 & 81.3\% \\
\bottomrule
\end{tabular}
\end{table}

The relative reduction is consistent even though the starting opportunity differs substantially.
Dolly and Alpaca have the largest decreases in the modeled straggler ratio
$\rho$. UltraChat begins much closer to balance and has a smaller decrease in
$\rho$ points despite a similar proportional reduction in $\rho-1$. CodeParrot and OpenWebMath have long uncapped tails. The 1,024-token cap limits their
modeled costs; BCL then reduces the remaining assignment variation in the simulation.

\section{NeMo-RL Streaming Evaluation}
\label{sec:nemo-evaluation}

\subsection{Setup and controls}

We fine-tune Meta's pretrained Llama-3.2-1B checkpoint
\citep{meta2024llama32} with NeMo-RL v0.7.0 SFT
\citep{nemorl} on one eight-H200 node (FSDP2, bf16, FlashAttention~3
\citep{shah2024flashattention3}). The
13 public corpora are Alpaca~\citep{taori2023alpaca}, Dolly~\citep{conover2023dolly},
OASST1~\citep{kopf2023openassistant}, OASST2~\citep{openassistant2023oasst2},
OpenHermes~\citep{teknium2023openhermes}, ShareGPT~\citep{sharegpt},
ShareGPT-GPT4~\citep{sharegpt_gpt4}, UltraChat~\citep{ding2023ultrachat},
CodeParrot~\citep{codeparrot}, FineWeb~\citep{penedo2024fineweb},
OpenWebMath~\citep{paster2023openwebmath},
OpenWebText~\citep{gokaslan2019openwebtext}, and
SlimPajama~\citep{soboleva2023slimpajama}. Each corpus is processed by NeMo's SFT
preprocessing with the same per-document p99 length cap in every arm. The
study-specific overlength truncation is shared across arms.

For the per-rank experiment, each of the eight NeMo-RL workers
owns eight of 64 Parquet shards. Two Ray readers per worker \citep{moritz2018ray}
tokenize that rank's records and schedule $K B$-document windows. A collector
inside the worker takes one pack per step and builds one unpadded row of $B$
documents for NeMo's training worker. No documents or scheduling metadata
pass between ranks or through the driver. BCL uses $K=32$, capped
squared-length cost, LPT initialization, two swap passes, and a seeded
pack shuffle;
default packing of input documents\footnote{This arrival-order control is
implemented with BCL's one-pack window ($K=1$).} uses the same per-rank
pipeline without lookahead. Both use the same model, loss, optimizer,
global batch size, FSDP2 training path, and train microbatch size of one.
Both disable per-step CUDA cache clearing and freeze Python garbage
collection after setup to avoid asymmetric allocation stalls.

\begin{table}[t]
\caption{NeMo-RL SFT comparisons. Only the first two arms share the per-rank
input path. MFFD is a pooled reference from the preceding campaign;
the default path was rerun in the current campaign.}
\label{tab:nemo-arms}
\centering
\scriptsize
\setlength{\tabcolsep}{3pt}
\begin{tabular}{p{0.31\columnwidth}p{0.60\columnwidth}}
\toprule
Arm & Assignment \\
\midrule
Default packing & Per-rank reader and collector; next $B$ documents from each rank, no lookahead. \\
BCL & Same per-rank path with $K=32$; exactly $B$ documents per rank. \\
Pooled MFFD & Earlier global-input path; per-step token-budget packing into one row per rank. \\
NeMo default path & Own loader and MFFD packer; potentially multiple rows per rank. \\
\bottomrule
\end{tabular}
\end{table}

\subsection{Measured step time}

Each arm has three 300-step repeats, excluding the first 20 steps.
We time the wall-clock interval between consecutive batch fetches, covering
input and training in both paths. NeMo's built-in step timer includes
per-rank fetch but excludes the earlier driver's fetch.
Table~\ref{tab:nemo-stream} reports the percentage speedups.
Against pooled MFFD, rank-local BCL shows $+5.0\%$ to $+16.7\%$ speedup across all
13 corpora. This compares complete input paths,
not packing algorithms in isolation. On the same per-rank path, $K=32$ shows
$+0.4\%$ to $+30.5\%$ mean speedup over default packing on 11 corpora,
isolating lookahead. The $K=1$ arm was not rerun
on CodeParrot or OpenWebMath at $B=8$ after earlier oversized-row OOMs.
NeMo's default loader and packer were rerun as a separate reference.

\begin{table*}[t]
\caption{NeMo-RL SFT wall-clock speedup at $B=8$, defined as
$100(t_{\mathrm{baseline}}/t_{\mathrm{BCL}}-1)$. Values are percentages;
$\pm$ terms are standard deviations in percentage points over three repeats.
Pooled MFFD uses a cross-campaign ratio of mean per-step times.}
\label{tab:nemo-stream}
\centering
\small
\begin{tabular}{lrrr}
\toprule
Corpus & vs.\ pooled MFFD (\%)\textsuperscript{*} & vs.\ default packing (\%) & vs.\ NeMo default (\%)\textsuperscript{\dag} \\
\midrule
Alpaca & $+5.0$ & $+0.4\pm1.4$ & $+122.8\pm2.5$ \\
Dolly & $+12.3$ & $+4.3\pm0.4$ & $+77.1\pm2.0$ \\
OpenHermes & $+11.3$ & $+7.5\pm0.5$ & $+74.8\pm0.8$ \\
OASST1 & $+9.5$ & $+4.9\pm1.2$ & $+109.1\pm1.0$ \\
OASST2 & $+11.1$ & $+5.5\pm0.3$ & $+104.2\pm2.1$ \\
FineWeb & $+16.7$ & $+22.7\pm0.2$ & $+58.5\pm0.9$ \\
SlimPajama & $+14.1$ & $+30.5\pm1.4$ & $+39.4\pm0.5$ \\
OpenWebText & $+15.6$ & $+30.2\pm0.1$ & $+55.8\pm1.3$ \\
UltraChat & $+10.4$ & $+12.5\pm0.9$ & $+74.7\pm1.9$ \\
ShareGPT-GPT4 & $+13.7$ & $+16.3\pm0.9$ & $+71.0\pm1.2$ \\
ShareGPT & $+13.3$ & $+16.4\pm0.9$ & $+87.1\pm1.7$ \\
OpenWebMath & $+15.2$ & -- & $+35.6\pm0.6$ \\
CodeParrot & $+13.0$ & -- & $+34.3\pm1.0$ \\
\bottomrule
\end{tabular}
\par\smallskip
\parbox{\textwidth}{\footnotesize\textsuperscript{*} Pooled MFFD uses the
earlier global-input path; this wall-clock comparison includes input-path changes
and is not paired by repeat. \textsuperscript{\dag} The NeMo default path uses
its own loader and may execute multiple rows per rank; its speedup is not an
isolated scheduling effect. Dashes indicate that the $K=1$ arm was not rerun
after earlier B=8 OOMs.}
\end{table*}

Reruns of NeMo's default path differ by at most 1.5\% between
campaigns, providing a check on node drift. Moving the same BCL arm from
driver-merged to per-rank input gives $+2.8\%$ to $+7.5\%$ speedup on 12
corpora ($+0.2\%$ on Alpaca). Per-step audits confirm one unpadded $B$-document row
per rank and matching loader and trainer token counts. Two readers per rank
usually keep input ready after warm-up.

\section{Multi-Node H200 Evaluation}
\label{sec:multinode-evaluation}

\subsection{Workload and reporting boundary}

The final evaluation asks whether the scheduling benefit persists from one to four nodes on an
internal sequence workload associated with the Feed Sequential Recommender training setting
described by \citet{hertel2026feed}. The training records are capped at 1,024 tokens before
scheduling. We report only length statistics and aggregate performance; no member content, raw
records, production model code, or execution harness is included.

The retained full characterization covers 7,187 shards and 320,137,728 records. Mean capped
length is 215.6 tokens, and 10.44\% of records lie at the cap. A separately retained 16-shard
sample contains 327,680 records, with mean 213.08, median 28, and 10.29\% at
the cap. The sample has similar mean length and cap rate.

\subsection{Length-sample scheduler probe}

Before measuring the model, a length-only probe applies the same fixed-cardinality scheduler to
the 16-shard sample at per-rank $B=384$. Table~\ref{tab:feed-sample-balance} shows the modeled
straggler ratio at eight and sixteen ranks. Moving from $K=1$ to $K=32$ removes approximately
82\% of straggler excess at both scales; $K=64$ provides a smaller additional reduction.

\begin{table}[H]
\caption{Modeled straggler ratios from scheduling 327,680 measured capped lengths,
not GPU step times.}
\label{tab:feed-sample-balance}
\centering
\small
\begin{tabular}{lrrr}
\toprule
Ranks & $K=1$ & $K=32$ & $K=64$ \\
\midrule
8 & 1.1438 & 1.0261 & 1.0192 \\
16 & 1.1931 & 1.0363 & 1.0286 \\
\bottomrule
\end{tabular}
\end{table}

The higher arrival-order ratio at 16 ranks is consistent with the rank-count effect in
Equation~\ref{eq:random-straggler}. BCL keeps the residual close to one without moving samples
between ranks.

\subsection{Compiled H200 setup}

The hardware comparison uses compiled execution \citep{ansel2024pytorch2} on H200 GPUs,
per-rank batch size 384, and BCL
window $K=32$. One, two, and four nodes correspond to 8, 16, and 32 ranks and derived global
batch sizes of 3,072, 6,144, and 12,288 records. The baseline is the same packed execution path
with BCL disabled. Reported values are post-warm-up seconds per step, aggregated across repeated
runs.
Global batch size rises with rank count; this is weak scaling at fixed per-rank batch size.

The comparison includes three BCL and four Packed repeats at 8 and
16 ranks, and three per arm at 32 ranks.

\subsection{Compiled step-time scaling}

Table~\ref{tab:h200-scaling} reports the primary measured comparison.
BCL is faster than packed execution at every topology. Step time falls by
16.45\%, 19.51\%, and 25.52\% at one, two, and four nodes, corresponding to
1.197$\times$, 1.242$\times$, and 1.343$\times$ speedups.
This rise is consistent with the larger straggler effect expected as more ranks
synchronize (Equation~\ref{eq:random-straggler}).

\begin{table}[H]
\caption{Primary measured compiled H200 step time. Values are mean $\pm$
sample standard deviation in seconds per step across repeated runs.}
\label{tab:h200-scaling}
\centering
\scriptsize
\setlength{\tabcolsep}{3pt}
\begin{tabular}{lrrr}
\toprule
Nodes / ranks & Packed & BCL & Speedup \\
\midrule
1 / 8 & $0.3915\pm0.0035$ & $0.3271\pm0.0005$ & 1.197$\times$ \\
2 / 16 & $0.4398\pm0.0035$ & $0.3540\pm0.0005$ & 1.242$\times$ \\
4 / 32 & $0.5436\pm0.0233$ & $0.4049\pm0.0034$ & 1.343$\times$ \\
\bottomrule
\end{tabular}
\end{table}

BCL reorders records only within bounded rank-local windows rather than globally
sorting by length, which can impose a training curriculum \citep{bengio2009curriculum}.
In Feed sequence-model evaluations, mean AUC differed from the reference by
just 0.07\% on average, with no observed loss in this aggregate metric.

\subsection{Peak-memory evidence}

Single-node telemetry at $B=384$ provides a supporting memory result. The busiest-rank peak is
$138.7\pm0.8$ GiB for packed execution and $124.9\pm4.2$ GiB with BCL, approximately 13.8 GiB
or 9.9\% lower.

\subsection{Calibrated sensitivity analysis}

Calibrated length-distribution replays extend the $K$ and $B$ sweeps beyond the measured
H200 setup. Table~\ref{tab:feed-calibrated} shows the trends. At $B=384$, the modeled straggler
ratio falls rapidly through $K=16$ and reaches 1.027 at $K=32$; the change to $K=64$ is smaller.
At fixed $K=32$, larger packs move closer to the calibrated global assignment target because
they average more per-record variation.

\begin{table}[H]
\caption{Calibrated Feed workload sensitivity. These values are model/replay outputs, not direct
hardware measurements.}
\label{tab:feed-calibrated}
\centering
\scriptsize
{\setlength{\tabcolsep}{2.2pt}
\begin{tabular}{lrrrrrrr}
\toprule
$K$ & 1 & 2 & 4 & 8 & 16 & 32 & 64 \\
\midrule
$\rho$ & 1.159 & 1.107 & 1.076 & 1.054 & 1.038 & 1.027 & 1.019 \\
\bottomrule
\end{tabular}}

\vspace{0.4em}

\begin{tabular}{lrrrrr}
\toprule
$B$ at $K=32$ & 32 & 96 & 256 & 384 & 1,024 \\
\midrule
\% modeled target & 94.5 & 96.7 & 98.0 & 98.3 & 98.9 \\
\bottomrule
\end{tabular}
\end{table}

From $K=32$ and $B=384$, increasing either parameter adds little in the replays.
The reported percentages measure proximity to a modeled assignment target, not
measured H200 speedup.

\section{Related Work}
\label{sec:related-work}

\newcommand{\relatedworklead}[1]{\paragraph{{\fontencoding{OT1}\fontfamily{cmr}\bfseries\selectfont #1}}}

\relatedworklead{Batching, packing, and padding elimination.}
Token-budget and length-aware batching reduce variation in the amount of padding or tokens
processed per batch, but may vary record cardinality and therefore require explicit normalization
and optimization policies \citep{ott2018scaling}. Sequence packing and variable-length attention
kernels remove computation on padding within a packed batch
\citep{krell2021packing,dao2022flashattention}. BCL is complementary: it preserves record
cardinality and changes only which records share packs that meet at a synchronization barrier.

\relatedworklead{Variable-length distributed training.}
Hydraulis, first released in 2024, jointly selects heterogeneous parallel strategies and assigns
data within and across model replicas \citep{li2026hydraulis}. WLB-LLM combines workload-aware
document packing with per-document context-parallel sharding \citep{wang2025wlbllm}. DynaPipe
builds variable-length microbatches and dynamically schedules pipeline execution and
communication \citep{jiang2024dynapipe}. These systems demonstrate that length-aware execution can recover
substantial throughput, but they alter parallelism, microbatch structure, or cross-worker data
assignment. BCL instead fixes the existing parallel plan and collective structure, and exposes a
rank-local data-loader option with exact record cardinality.

\relatedworklead{Scheduling foundations.}
The assignment core is not a new scheduling heuristic. Its initialization follows classical
longest-processing-time list scheduling \citep{graham1969bounds}; the NeMo comparison uses
modified first-fit decreasing \citep{johnson1985mffd} as a per-step reference.
BCL's systems contribution is the bounded, streaming deployment
contract and its coordination boundary. The motivation follows the broader synchronous-straggler
literature: each step waits for its slowest worker, and tail effects become more consequential as
more workers participate \citep{li2020pytorch,chen2016revisiting,dean2013tail}.

\relatedworklead{Recent approaches.}
Several recent systems address related workload skew through different forms of coordination.
Libra uses a bounded sequence pool to address long-context attention-workload skew
\citep{wang2026libra}; ODC revisits parameter-server-style decoupling for LLM post-training
\citep{wan2026odc}; and RoutePack co-plans mixture-of-experts placement with attention-aware data
packing over an optimizer-step window \citep{shen2026routepack}. RoutePack plans globally for a mixture-of-experts workload, while BCL reorders
fixed-size packs within independent reader windows. If rank-window totals differ
substantially, coordinated assignment may be needed (Section~\ref{sec:boundaries}).

\section{Deployment Boundaries}
\label{sec:discussion}

Before scheduling, render and tokenize records as training will consume them and
choose a length cap and batch size that fit in memory. Earlier probes used raw
rendered-length $\mathrm{CV}\geq2$ to trigger proactive p99 capping; lower-CV
corpora also OOMed uncapped, so CV is a heuristic rather than a feasibility test.
With feasible records, a length-only sweep can suggest a lookahead window, but
its benefit and staging cost must be measured on the target trainer. A dominant
record needs a shorter cap or a different representation; persistently heavier
rank windows call for resharding or coordination. The NeMo comparison holds
p99 caps fixed and does not evaluate their effect on training quality.

BCL changes which records share an optimizer step. Workloads with ordering
constraints must define legal permutations before applying it. An exact
mid-stream restart would need to restore reader positions, staged packs,
shuffle state, and the per-rank reader merge. The current NeMo experiment
does not test exact restart or held-out quality under that reordering.

\section{Conclusion}
\label{sec:conclusion}

Equal record counts do not guarantee equal work in variable-length distributed training.
BCL groups each reader's bounded window into balanced fixed-cardinality packs without
exchanging cost metadata between readers or changing the model and collectives.

Paired simulation over six public length distributions reduces per-rank modeled
pack-cost variability by
81.4--82.8\%. In NeMo-RL SFT, rank-local BCL improves wall-clock step time
by 1.050--1.167$\times$ over pooled MFFD across 13 corpora, including
input-path differences. Against matched no-lookahead packing, BCL is
1.004--1.305$\times$ faster on 11 corpora and completes the other two,
where earlier $B=8$ no-lookahead runs OOMed. On the internal H200
workload, compiled measurements show 1.197--1.343$\times$ speedup from one
to four nodes.

BCL is most useful when records fit and the reader windows have similar total costs.
Outlying records need capping or a change in representation; consistently heavier
windows call for resharding or coordination.

\section*{Acknowledgments}
We thank our leaders Animesh Singh and Dave Burgess for their invaluable expertise in the
ML infrastructure stack and open-source strategy.

\section*{Data and Model Use}
The public-dataset fine-tuning experiments in Section 6 use publicly available datasets solely for research and benchmarking, to compare BCL against existing open-source frameworks. The fine-tuned Llama-3.2-1B models produced in these experiments, and the datasets as used in this work, are not used in LinkedIn production systems.

\bibliography{references}
\bibliographystyle{mlsys2026}

\end{document}